\documentclass[aps,prd,nofootinbib,superscriptaddress,onecolumn,notitlepage,preprintnumbers]{revtex4-1}

\usepackage{amsmath}
\usepackage{amssymb}
\usepackage{graphicx,epsfig,wrapfig,bbm}
\usepackage{epstopdf}
\usepackage{bm}
\usepackage{placeins}
\usepackage{float}
\usepackage[normalem]{ulem} 
\usepackage{color}

\usepackage[utf8]{inputenc}
\usepackage{braket}
\usepackage{slashed}
\usepackage[usenames,dvipsnames]{xcolor}
\usepackage[english]{babel}
\usepackage{verbatim}
\usepackage{empheq}
\usepackage[pdftex]{hyperref}

\def\nn{\nonumber}
\def\cd{{\cdot}}

\usepackage[normalem]{ulem} 
\renewcommand\sout{\bgroup \color[rgb]{0.55,0.00,0.99} \ULdepth=-.5ex \ULset}

\newcommand{\st}{{\scriptscriptstyle T}}
\DeclareMathOperator{\tr}{Tr}

\allowdisplaybreaks

\begin{document}

\title{Positivity bounds on gluon TMDs for hadrons of spin $\le$ 1}

\newcommand*{\Nikhef}{Nikhef, Science Park 105, NL-1098 XG Amsterdam, The Netherlands}\affiliation{\Nikhef}
\newcommand*{\Uamsterdam}{Department of Physics and Astronomy, VU University Amsterdam, De Boelelaan 1081, NL-1081 HV Amsterdam, The Netherlands}\affiliation{\Uamsterdam}
\preprint{Nikhef 2017-041}

\author{Sabrina Cotogno}\email{scotogno@nikhef.nl}\affiliation{\Nikhef}\affiliation{\Uamsterdam}
\author{Tom van Daal}\email{tvdaal@nikhef.nl}\affiliation{\Nikhef}\affiliation{\Uamsterdam}
\author{Piet J. Mulders}\email{pietm@nikhef.nl}\affiliation{\Nikhef}\affiliation{\Uamsterdam}


\begin{abstract}
We consider the transverse momentum dependent gluon distribution functions (called gluon TMDs) by studying the light-front gluon-gluon correlator, extending the results for unpolarized and vector polarized targets to also include tensor polarized targets -- the latter type of polarization is relevant for targets of spin $\ge1$. The light-front correlator includes process-dependent gauge links to guarantee color gauge invariance. As from the experimental side the gluon TMDs are largely unknown, we present positivity bounds for combinations of leading-twist gluon distributions that may be used to estimate their maximal contribution to observables. Since the gluonic content of hadrons is particularly relevant in the small-$x$ kinematic region, we also study these bounds in the small-$x$ limit for the dipole-type gauge link structure using matrix elements of a single Wilson loop. 
\end{abstract}

\maketitle

\section{Introduction} \label{intro}
Understanding the structure of hadrons in terms of their elementary degrees of freedom is a challenging task far from being fulfilled. Whereas we have acquired considerable knowledge on the distributions in momentum and coordinate space of quarks and gluons inside nucleons, very little has been studied about the fundamental constituents of hadrons of spin higher than $1/2$.

In 1988 it was pointed out that in high-energy processes involving spin-$1$ hadrons one can define collinear quark structure functions (called $b_{1,2,3,4}$) that can be measured in tensor polarized targets~\cite{Hoodbhoy:1988am}. The simplest and hence most-studied (nuclear) spin-$1$ system is the deuteron. The extraction of the function $b_1$ for the deuteron was performed by the HERMES collaboration in 2005~\cite{Airapetian:2005cb}. The data collected and the parametrization proposed~\cite{Kumano:2010vz} deviate significantly from the standard theoretical predictions~\cite{Hoodbhoy:1988am,Khan:1991qk,Miller:2013hla,Cosyn:2017fbo}, both for the $x$ behavior and the magnitude, although the experimental uncertainties leave room for improvements. This suggests that, for the deuteron, dynamics beyond quarks and gluons confined within the individual nucleons is needed to describe it. More measurements of $b_1$ will be performed as part of the 12 GeV program at Jefferson Lab (JLab)~\cite{Slifer:2014bda}. Experimental information on spin-1 hadrons such as (virtual) $\rho$ mesons would allow us to thoroughly study such different quark contributions and dynamics, as recently explored with model calculations in~\cite{Ninomiya:2017ggn}; this is currently beyond experimental reach.

Another interesting and even less investigated aspect of hadrons of spin $\ge 1$ is the \emph{gluonic} structure linked to tensor polarization of the target. More knowledge on gluon distributions could yield new insights into the internal dynamics of nuclei. A collinear structure function for gluons in spin-$1$ targets, called $\Delta(x)$, was first defined in~\cite{Jaffe:1989xy}. The authors  pointed out that this observable is related to a transfer of two units of helicity to the nuclear target, and vanishes for any target of spin smaller than $1$. They recognized that there must exist a tower of gluon operators contributing to the scattering amplitude with such a double-helicity flip that cannot be linked to single nucleons; rather, they are exclusive to hadrons and nuclei of spin $\ge 1$. In the parton model language, $\Delta(x)$ describes linearly polarized gluons in targets with transverse tensor polarization. Aspects of this function (its first moment and a positivity bound) have recently been studied on the lattice in~\cite{Detmold:2016gpy}, and also experimental interest has been shown. An example of the latter is a letter of intent about the extraction of $\Delta(x)$ that has been presented at JLab with the aim of investigating the region of small $x$ using nitrogen targets~\cite{DetmoldJLab}. On the other hand, the extraction of gluon functions remains challenging. In this respect, the construction of the Electron-Ion Collider (EIC)~\cite{Accardi:2012qut} is very promising: it could unravel the gluon content of hadrons by measuring a wide variety of gluon observables in a region where they are not overwhelmed by quark observables (as is the case at present facilities).

So far we have discussed quantities that only depend on the partonic momentum collinear  to the direction of motion of the parent hadron. Going beyond the collinear case, one can define transverse momentum dependent (TMD) parton distribution functions (PDFs) (or simply called TMDs). These TMDs appear in the parametrization of a TMD correlator, which is a bilocal matrix element containing nonlocal field strength operators and Wilson lines (also called gauge links). The latter are necessary to guarantee color gauge invariance by bridging the nonlocality, and give rise to a process dependence of the TMDs. In the simplest case, the gauge link structure is just built from future- and past-pointing staple-like gauge links. The TMDs occurring in the description of spin-$1$ hadrons have been systematically defined for both quarks~\cite{Bacchetta:2000jk} and gluons~\cite{Boer:2016xqr}. In this paper we continue the study started in~\cite{Boer:2016xqr} on the properties of and the relations between the gluon TMDs for spin-$1$ hadrons. More specifically, we will derive positivity bounds, i.e. model-independent inequalities, that help relating and estimating the magnitude of the gluon TMDs about which very little, or almost nothing, is currently known. An analogous analysis for spin-$1/2$ hadrons was performed in~\cite{Mulders:2000sh}, and the quark case has already been considered for both spin-$1/2$ and spin-$1$ hadrons~\cite{Bacchetta:1999kz,Bacchetta:2001rb}. Bounds on collinear gluon functions for spin-$1$ hadrons were recently presented in~\cite{Cotogno:2017mwy}, and we will include them here for completeness.

It was recently shown in~\cite{Boer:2015pni,Boer:2016xqr} that for the so-called dipole-type gauge link structure, the number of independent gluon TMDs greatly reduces in the small-$x$ limit. Hence, the description of gluon TMDs simplifies significantly in the kinematic region where they are expected to be most important. The dipole-type TMD correlator has one future- and one past-pointing gauge link and can, in the limit of small $x$, be related to a correlator containing a single Wilson loop. The latter correlator can be parametrized in terms of TMDs for which we will also derive bounds. 

The outline of this work is as follows: in section~\ref{sec:2} we briefly recall the parametrization of the gluon-gluon TMD correlator for spin-$1$ hadrons in terms of gluon TMDs. In section~\ref{sec:3} we present a set of inequalities for those TMDs, as well as for their collinear counterparts. For completeness, we will also discuss the spin-$1/2$ case. Subsequently, in section~\ref{sec:4} we derive bounds for the gluon TMDs that apply to the small-$x$ kinematic region, both for spin-$1$ and spin-$1/2$ hadrons. Finally, we discuss our conclusions in section~\ref{sec:5}.

\section{The gluon-gluon correlation function} \label{sec:2}
Recently in~\cite{Boer:2016xqr} the gluon-gluon TMD (or light-front) correlator was systematically parametrized in terms of TMDs considering unpolarized, vector, as well as tensor polarized targets. The latter type of polarization is needed for the description of hadrons of spin $\geq 1$. In this section we briefly summarize the results of this parametrization. 

We use a decomposition for the gluon momentum $k$ in terms of the hadron momentum $P$ and the lightlike four-vector $n$, such that 
\begin{equation}
    k^\mu = x P^\mu + k_\st^\mu + (k\cd P - xM^2) \,n^\mu ,
\end{equation}
satisfying $P \cd n = 1$ and $P^2 = M^2$, where $M$ is the mass of the hadron. The gluon-gluon TMD correlator for spin-$1$ hadrons is defined as:
\begin{equation}
    \Gamma^{[U,U^\prime]\,\mu\nu;\rho\sigma}(x,\bm{k}_\st,P,n;S,T) \equiv \int \left. \frac{d\xi\cd P\, d^2\bm{\xi}_\st}{(2\pi)^3} \;e^{ik\cdot\xi} \bra{P;S,T} \tr_c \left(F^{\mu\nu}(0) \,U_{[0,\xi]}^{\phantom{\prime}} F^{\rho\sigma}(\xi) \,U_{[\xi,0]}^\prime \right) \ket{P;S,T} \right|_{\xi\cd n=0} ,
    \label{e:TMDcorrelator}
\end{equation}
where the process-dependent Wilson lines $U_{[0,\xi]}^{\phantom{\prime}}$ and $U_{[\xi,0]}^\prime$ are needed to ensure color gauge invariance. We will consider specific gauge link structures built from the future- and past-pointing Wilson lines $U_{[0,\xi]}^{[+]}$ and $U_{[0,\xi]}^{[-]}$ respectively, defined as 
\begin{equation}
    U_{[0,\xi]}^{[\pm]} \equiv U^n_{[0,\bm{0}_\st;\pm\infty,\bm{0}_\st]} \,U^T_{[\pm\infty,\bm{0}_\st;\pm\infty,\bm{\xi}_\st]} \,U^n_{[\pm\infty,\bm{\xi}_\st;\xi \cdot P,\bm{\xi}_\st]} ,
    \label{e:link}
\end{equation}
where the pieces denoted by $U^n$ are links along the direction of $n$, and the transverse piece $U^T$ is located at (plus or minus) light cone infinity. Counting powers of the inverse hard scale relevant in the process, leads to the definition of the leading-power (usually referred to as leading twist or twist-2) correlator,
\begin{equation}
    \Gamma^{ij}(x,\bm{k}_\st;S,T) \equiv \Gamma^{[U,U^\prime]\,ni;nj}(x,\bm{k}_\st,P,n;S,T) ,
    \label{leadingcorrelator}
\end{equation}
where the superscripts $n$ on the right-hand side denote Lorentz contractions with the vector $n$.

The correlator in eq.~\eqref{leadingcorrelator} has been averaged over the target spin states and is defined as
\begin{equation}
    \Gamma^{ij}(x,\bm{k}_\st;S,T) \equiv \tr \left( \rho(S,T) \,G^{ij}(x,\bm{k}_\st) \right) ,
    \label{e:average}
\end{equation}
where the information on the spin states of the parent hadron is encoded in the $3\times3$ density matrix $\rho(S,T)$ and the combined information on the hadron and gluon spins is contained in $G^{ij}(x,\bm{k}_\st)$. The density matrix is parametrized in terms of a spacelike spin vector $S$ and a symmetric traceless spin tensor $T$ (for a spin-$1/2$ hadron only $S$ is needed). Ensuring the relations $P\cd S=0$ and $P_\mu T^{\mu\nu}=0$, they can be parametrized in terms of $P$ and $n$ as follows~\cite{Leader:2001gr,Bacchetta:2000jk,Bacchetta:2002}:
\begin{align}
    S^\mu &= S_L \frac{P^\mu}{M} + S_T^\mu - MS_L \,n^\mu , \label{e:S} \\
    T^{\mu\nu} &= \frac{1}{2} \left[ \frac{2}{3} S_{LL} \,g_\st^{\mu\nu} + \frac{4}{3} S_{LL} \frac{P^\mu P^\nu}{M^2} + \frac{S_{LT}^{\{\mu}P^{\nu\}}}{M} + S_{TT}^{\mu\nu} \right. \nn \\
    &\left. \qquad\;\; - \,\frac{4}{3} S_{LL} P^{\{\mu}n^{\nu\}} - M S_{LT}^{\{\mu}n^{\nu\}} + \frac{4}{3} M^2 S_{LL} \,n^\mu n^\nu \vphantom{\frac{P^\mu P^\nu}{M^2}} \right] , \label{e:T}
\end{align}
where we have used the metric tensor in transverse space defined as $g_\st^{\mu\nu} \equiv g^{\mu\nu} - P^{\{\mu} n^{\nu\}}$ (curly brackets denote symmetrization of the indices), with nonvanishing elements $g_\st^{11} = g_\st^{22} = -1$. Choosing a Cartesian basis of $3\times3$ matrices consisting of the identity matrix $I$, the three-dimensional (generalized) Pauli matrices $\Sigma^i$, and the five bilinear combinations $\Sigma^{ij} \equiv \tfrac{1}{2} (\Sigma^i \Sigma^j + \Sigma^j \Sigma^i) - \tfrac{2}{3} I \delta^{ij}$, and making use of eqs.~\eqref{e:S} and~\eqref{e:T}, the density matrix takes the form:
\begin{equation}
    \rho(S,T) = \frac{1}{3} \left( I + \frac{3}{2} S^i \Sigma^i + 3 \,T^{ij} \Sigma^{ij} \right) = \left( \begin{array}{ccc}
    \frac{1}{3} + \frac{S_L}{2} + \frac{S_{LL}}{3} & \frac{S_T^x - iS_T^y}{2\sqrt{2}} + \frac{S_{LT}^x - iS_{LT}^y}{2\sqrt{2}} &
    \frac{S_{TT}^{xx} - iS_{TT}^{xy}}{2} \\[5pt]
    \frac{S_T^x + iS_T^y}{2\sqrt{2}} + \frac{S_{LT}^x + iS_{LT}^y}{2\sqrt{2}} & \frac{1}{3} - \frac{2 S_{LL}}{3} & \frac{S_T^x - iS_T^y}{2\sqrt{2}} - \frac{S_{LT}^x - iS_{LT}^y}{2\sqrt{2}} \\[5pt]
    \frac{S_{TT}^{xx} + iS_{TT}^{xy}}{2} & \frac{S_T^x + iS_T^y}{2\sqrt{2}} - \frac{S_{LT}^x + iS_{LT}^y}{2\sqrt{2}} & \frac{1}{3} - \frac{S_L}{2} + \frac{S_{LL}}{3} \\[5 pt]
    \end{array} \right) .
    \label{e:density}
\end{equation}
In the rest of this paper the dependence of the correlators on $S$ and $T$ will be implicit.

The correlator in eq.~\eqref{leadingcorrelator} has been parametrized in terms of TMDs employing symmetric traceless tensors $k_\st^{i_1 \ldots i_n}$ built from the partonic transverse momentum $k_\st$ (see appendix C of~\cite{Boer:2016xqr} for the relevant definitions up to rank $n=4$). The use of those tensor structures ensures that the TMDs occurring in the parametrization are \emph{of definite rank}. This has the advantage that there is a one-to-one correspondence between the functions in momentum space ($k_\st$-space) and in coordinate space ($b_\st$-space, where $b_\st$ is Fourier conjugate to $k_\st$), which is an important feature when considering TMD evolution equations~\cite{Boer:2016xqr,Signori:2016lvd,vanDaal:2016glj}. Separating the various possible hadronic polarization states, the correlator in eq.~\eqref{leadingcorrelator} can be parametrized in terms of leading-twist gluon TMDs of definite rank as follows:~\cite{Boer:2016xqr} 
\begin{equation}
    \Gamma^{ij}(x,\bm{k}_\st) = \Gamma_U^{ij}(x,\bm{k}_\st) + \Gamma_L^{ij}(x,\bm{k}_\st) + \Gamma_T^{ij}(x,\bm{k}_\st) + \Gamma_{LL}^{ij}(x,\bm{k}_\st) + \Gamma_{LT}^{ij}(x,\bm{k}_\st) + \Gamma_{TT}^{ij}(x,\bm{k}_\st) ,
    \label{e:gamma_vp_tmds}
\end{equation}
where:\footnote{We define the transverse four-vector $a_\st^\mu$ to have light cone components $(0,0,\bm{a}_\st)$, where $\bm{a}_\st$ is a two-dimensional vector on the transverse plane. This implies e.g.\ that $a_\st^2 = -\bm{a}_\st^2$.}
\begin{align}
    \Gamma_U^{ij}(x,\bm{k}_\st) &= \frac{x}{2} \left[ - \,g_\st^{ij} \,f_1(x,\bm{k}_\st^2) + \frac{k_\st^{ij}}{M^2} \,h_1^{\bot}(x,\bm{k}_\st^2) \right] , \label{e:term1} \\
    \Gamma_L^{ij}(x,\bm{k}_\st) &= \frac{x}{2} \left[ i \epsilon_\st^{ij} S_L \,g_1(x,\bm{k}_\st^2) + \frac{{\epsilon_\st^{\{i}}_\alpha k_\st^{j\}\alpha} S_L}{2M^2} \,h_{1L}^\perp(x,\bm{k}_\st^2) \right] , \\
    \Gamma_T^{ij}(x,\bm{k}_\st) &= \frac{x}{2} \left[ - \,\frac{g_\st^{ij} \epsilon_\st^{S_\st k_\st}}{M} \,f_{1T}^\perp(x,\bm{k}_\st^2) + \frac{i \epsilon_\st^{ij} \bm{k}_\st \cd \bm{S}_\st}{M} \,g_{1T}(x,\bm{k}_\st^2) \right. \nn \\
    &\qquad\quad\!\! \left. - \,\frac{\epsilon_\st^{k_\st\{i} S_\st^{j\}} + \epsilon_\st^{S_\st\{i} k_\st^{j\}}}{4M} \,h_1(x,\bm{k}_\st^2) - \frac{{\epsilon_\st^{\{i}}_\alpha k_\st^{j\}\alpha S_\st}}{2M^3} \,h_{1T}^\perp(x,\bm{k}_\st^2) \right] , \label{e:term3} \\
    \Gamma_{LL}^{ij}(x,\bm{k}_\st) &= \frac{x}{2} \left[ - \,g_\st^{ij} S_{LL} \,f_{1LL}(x,\bm{k}_\st^2) + \frac{k_\st^{ij} S_{LL}}{M^2} \,h_{1LL}^{\bot}(x,\bm{k}_\st^2) \right] , \\
    \Gamma_{LT}^{ij}(x,\bm{k}_\st) &= \frac{x}{2} \left[ - \,\frac{g_\st^{ij} \bm{k}_\st \cd \bm{S}_{LT}}{M} \,f_{1LT}(x,\bm{k}_\st^2) 
    + \frac{i \epsilon_\st^{ij} \epsilon_\st^{S_{LT}k_\st}}{M} \,g_{1LT}(x,\bm{k}_\st^2) \right. \nn \\
    &\qquad\quad\!\! \left. + \,\frac{S_{LT}^{\{i} k_\st^{j\}}}{M} \,h_{1LT}(x,\bm{k}_\st^2) + \frac{k_\st^{ij\alpha} {S_{LT}}_\alpha}{M^3} \,h_{1LT}^{\bot}(x,\bm{k}_\st^2) \right] , \\
    \Gamma_{TT}^{ij}(x,\bm{k}_\st) &= \frac{x}{2} \left[ - \,\frac{g_\st^{ij} k_\st^{\alpha\beta} {S_{TT}}_{\alpha\beta}}{M^2} \,f_{1TT}(x,\bm{k}_\st^2) 
    + \frac{i \epsilon_\st^{ij} {\epsilon^{\beta}_\st}_\gamma k_\st^{\gamma\alpha} {S_{TT}}_{\alpha\beta}}{M^2} \,g_{1TT}(x,\bm{k}_\st^2) \right. \nn \\
    &\qquad\quad\!\! \left. + \,S_{TT}^{ij} \,h_{1TT}(x,\bm{k}_\st^2) + \frac{{S_{TT}^{\{i}}_\alpha k_\st^{j\}\alpha}}{M^2} \,h_{1TT}^{\bot}(x,\bm{k}_\st^2) + \frac{k_\st^{ij\alpha\beta} {S_{TT}}_{\alpha\beta}}{M^4} \,h_{1TT}^{\bot\bot}(x,\bm{k}_\st^2) \right] .
\end{align}
We have defined the transverse antisymmetric tensor as $\epsilon_\st^{\mu\nu} \equiv \epsilon^{Pn\mu\nu}$, with nonzero components $\epsilon_\st^{12} = -\epsilon_\st^{21} = 1$. Throughout the paper, the dependence of the TMDs on the gauge link structure is implicit.

Integrating the TMD correlator in eq.~\eqref{leadingcorrelator} over transverse momentum, we obtain the collinear correlator
\begin{equation}
    \Gamma^{ij}(x) \equiv \int \left. \frac{d\xi\cd P}{2\pi} \;e^{ik\cdot\xi} \bra{P;S,T} \tr_c \left(F^{ni}(0) \,U_{[0,\xi]}^{\phantom{\prime}} F^{nj}(\xi) \,U_{[\xi,0]}^\prime \right) \ket{P;S,T} \right|_{\xi\cd n=\bm{\xi}_\st=0} ,
    \label{e:PDFcorrelator}
\end{equation}
The parametrization of this correlator in terms of collinear PDFs is given by
\begin{equation}
    \Gamma^{ij}(x) = \frac{x}{2} \left[ - \,g_\st^{ij} \,f_1(x) + i \epsilon_\st^{ij} S_L \,g_1(x) - g_\st^{ij} S_{LL} \,f_{1LL}(x) + S_{TT}^{ij} \,h_{1TT}(x) \vphantom{\frac{x}{2}} \right] ,
    \label{e:PDFpar}
\end{equation}
where $f_1(x) \equiv \int d^2 \bm{k}_\st \,f_1(x,\bm{k}_\st^2)$, and similarly for the other functions. The collinear functions are universal as the gauge links are now unique straight lines along the direction of $n$. We note that $h_{1TT}(x)$ appears in the structure function $\Delta(x)$ defined in~\cite{Jaffe:1989xy} (the latter is called $\Delta_2G(x)$ in~\cite{Artru:1989zv}).

\section{Positivity bounds on gluon distributions} \label{sec:3}
As is well known, one can impose positivity constraints on the hadronic tensor and find a probabilistic interpretation for some of the distribution functions \cite{Artru:2008cp}. Positivity bounds on gluon TMDs were studied in~\cite{Mulders:2000sh} for spin-$1/2$ hadrons and, by applying the same strategy, we extend here this analysis to spin-$1$ hadrons. The starting point is the idea that the correlator $\Gamma$ can be seen as a $2\times2$ matrix in the two transverse gluon polarizations, given by $\Gamma^{ij} = \rho_{s^\prime s} \,G^{ij}_{s s^\prime}$ (see eq.~\eqref{e:average}), where $s,s^\prime$ label the hadronic polarization states. The quantity $G$ can be regarded as a $6\times6$ matrix in gluon $\otimes$ hadron spin space. As we will show explicitly, $G$ turns out to be positive semidefinite, a property which allows for setting constraints on the gluon distributions. First, we will derive bounds for the TMD case, and subsequently we will consider the transverse momentum integrated case. For completeness, we will also include the bounds that apply to spin-$1/2$ hadrons, completing the study of~\cite{Mulders:2000sh} where T-odd functions were not included.

\subsection{Bounds on transverse momentum dependent functions}
In this subsection we derive bounds for the gluons TMDs that appear in the parametrization given in eq.~\eqref{e:gamma_vp_tmds}. We choose the same basis for the matrix $G$ as in~\cite{Mulders:2000sh}, namely we use circular gluon polarizations, given by $\ket{\pm} = \mp \,\frac{1}{\sqrt 2} \left(\ket x \pm i \ket y\right)$. At leading twist, this matrix is given by
\begin{equation}
    G = \frac{x}{2} \left( \begin{array}{cc}
    A&B \\
    B^\dagger&C \\
    \end{array} \right) ,
    \label{Matrixsixbysix}
\end{equation}
where

\begin{equation}
\vspace{0.5cm}
\footnotesize
\setlength{\arraycolsep}{2.5pt} 
\medmuskip = 0.2mu
    A = \!\left( \begin{array}{ccc}
    {f_{1}}+\frac{{f_{1LL}}}{2}-{g_1} & \frac{e^{-i {\phi}} k }{\sqrt{2}
    M}\left({f_{1LT}}+i {f_{1T}^\bot}-{g_{1T}}-i {g_{1LT}}+{h_{1LT}}\right) & \frac{e^{-2 i {\phi}} k^2}{M^2}\left({f_{1TT}}+i {g_{1TT}}-{h_{1TT}^\bot}\right) \\[5pt]
    \frac{e^{i {\phi}} k}{\sqrt{2} M}\left({f_{1LT}}-i {f_{1T}^\bot}-{g_{1T}}+i {g_{1LT}}+{h_{1LT}}\right) & {f_{1}}-{f_{1LL}} & -\frac{e^{-i
    {\phi}} k }{\sqrt{2} M} \left({f_{1LT}}-i {f_{1T}^\bot}+{g_{1T}}-i {g_{1LT}}+{h_{1LT}}\right)\\[5pt]
    \frac{e^{2 i {\phi}} k^2 }{M^2} \left({f_{1TT}}-i {g_{1TT}}-{h_{1TT}^\bot}\right)& -\frac{e^{i {\phi}} k}{\sqrt{2} M}\left({f_{1LT}}+i {f_{1T}^\bot}+{g_{1T}}+i
    {g_{1LT}}+{h_{1LT}}\right) & {f_{1}}+\frac{{f_{1LL}}}{2}+{g_1}\nn \\
    \end{array} \right) ,
\end{equation}
\begin{equation}
\vspace{0.5cm}
\footnotesize
\setlength{\arraycolsep}{2.5pt} 
\medmuskip = 0.2mu
    B = \!\left( \begin{array}{ccc}
    -\frac{e^{-2 i {\phi}} k^2}{4 M^2} \left(2 {h_{1}^\bot} + {h_{1LL}^\bot}-2 i {h_{1L}^\bot}\right) & \frac{e^{-3 i {\phi}} k^3}{2
    \sqrt{2} M^3}\left({h_{1LT}^\bot}+i {h_{1T}^\bot}\right) & -\frac{e^{-4 i {\phi}} k^4}{2 M^4} h_{1TT}^{\bot\bot}\\[5pt]
    -\frac{e^{-i {\phi}} k}{\sqrt{2} M}\left(2{h_{1LT}} - i{h_{1}}\right) & -\frac{e^{-2 i {\phi}} k^2}{2 M^2}\left({h_{1}^\bot - {h_{1LL}^\bot}}\right) & -\frac{e^{-3 i {\phi}} k^3}{2 \sqrt{2} M^3}\left({h_{1LT}^\bot}-i {h_{1T}^\bot}\right) \\[5pt]
    -2 {h_{1TT}} & \frac{e^{-i {\phi}} k }{\sqrt{2} M}\left(2 {h_{1LT}}+i {h_{1}}\right) & -\frac{e^{-2 i {\phi}} k^2}{4 M^2}\left(2 {h_{1}^\bot + {h_{1LL}^\bot}+2 i{h_{1L}^\bot}}\right) \nn\\
    \end{array} \right) ,
\end{equation}
\begin{equation}
\vspace{0.5cm}
\footnotesize
\setlength{\arraycolsep}{2.5pt} 
\medmuskip = 0.2mu
    C = \!\left( \begin{array}{ccc}
    {f_{1}}+\frac{{f_{1LL}}}{2}+{g_1} & \frac{e^{-i {\phi}} k }{\sqrt{2}
    M}\left({f_{1LT}}+i {f_{1T}^\bot}+{g_{1T}}+i {g_{1LT}}+{h_{1LT}}\right) & \frac{e^{-2 i {\phi}} k^2}{M^2}\left({f_{1TT}}-i {g_{1TT}}-{h_{1TT}^\bot}\right) \\[5pt]
    \frac{e^{i {\phi}} k}{\sqrt{2} M}\left({f_{1LT}}-i {f_{1T}^\bot}+{g_{1T}}-i {g_{1LT}}+{h_{1LT}}\right) & {f_{1}}-{f_{1LL}} & -\frac{e^{-i
    {\phi}} k }{\sqrt{2} M} \left({f_{1LT}}-i {f_{1T}^\bot}-{g_{1T}}+i {g_{1LT}}+{h_{1LT}}\right)\\[5pt]
    \frac{e^{2 i {\phi}} k^2 }{M^2} \left({f_{1TT}}+i {g_{1TT}}-{h_{1TT}^\bot}\right)& -\frac{e^{i {\phi}} k}{\sqrt{2} M}\left({f_{1LT}}+i {f_{1T}^\bot}-{g_{1T}}-i
    {g_{1LT}}+{h_{1LT}}\right) & {f_{1}}+\frac{{f_{1LL}}}{2}-{g_1}\nn \\
    \end{array} \right) ,
\end{equation}
where for convenience we have suppressed the argument $(x,\bm{k}_\st^2)$ of the functions. Furthermore, we have expressed $\bm{k}_\st$ in terms of its polar coordinates $k$ and $\phi$. From symmetry considerations it follows that block $C$ is the parity transformed of $A$ and the off-diagonal blocks are Hermitian conjugates (see appendix A of~\cite{Boer:2016xqr} for more details on the parity, Hermiticity, time-reversal, and charge-conjugation properties of the gluon correlator).

To make more apparent the properties of the matrix $G$, we write its elements in the following form:
\begin{align}
    G^{ij}_{s s^\prime}(x,\bm{k}_\st) &\equiv \int \left. \frac{d\xi\cd P\, d^2\bm{\xi}_\st}{(2\pi)^3} \;e^{ik\cdot\xi} \bra{P;s} \tr_c \left( F^{ni}(0) \,F^{nj}(\xi) \right) \ket{P;s^\prime} \right|_{\xi\cd n=0} \nn \\
    &= \sum_m \,\tr_c \left( \bra{P_m} F^{ni}(0) \ket{P;s}^* \bra{P_m} F^{nj}(0) \ket{P;s^\prime} \right) \delta(P_m \cd n - (1-x)) \;\delta^{(2)}({\bm{P}_m}_\st + \bm{k}_\st) ,
    \label{relationTMD}
\end{align}
where we inserted a complete set a momentum eigenstates $\{\ket{P_m}\}$. We infer from eq.~\eqref{relationTMD} that, in any basis, the diagonal elements are given by absolute squares. In particular, it follows that the eigenvalues of $G$ in eq.~\eqref{Matrixsixbysix} must be $\geq 0$, or, equivalently, that $G$ is positive semidefinite. This can be used to set constraints on the TMDs. Given the limited amount of information we have on the gluon functions, we will refrain from diagonalizing the full $6\times6$ matrix and rather restrict ourselves to finding the eigenvalues of its $2\times2$ principal minors. Due to the symmetry properties of $G$, some minors yield the same bounds; we obtain the following nine inequalities:
\begin{align}
    \frac{\bm{k}_\st^2}{2M^2} \,|h_1^\perp - h_{1LL}^\perp| &\leq f_1 - f_{1LL} , \label{e:bound1} \\
    \frac{\bm{k}_\st^4}{16 M^4} \left[ 4 (h_{1L}^\perp)^2 + (2 h_1^\perp + h_{1LL}^\perp)^2 \right] &\leq \left( f_1 + \frac{f_{1LL}}{2} + g_1 \right) \left( f_1 + \frac{f_{1LL}}{2} - g_1 \right) , \\
    \frac{\bm{k}_\st^2}{2M^2} \left( h_1^2 + 4 h_{1LT}^2 \right) &\leq (f_1 - f_{1LL}) \left( f_1 + \frac{f_{1LL}}{2} + g_1 \right) , \\
    \frac{\bm{k}_\st^6}{8 M^6} \left[ (h_{1T}^\perp)^2 + (h_{1LT}^\perp)^2 \right] &\leq (f_1 - f_{1LL}) \left( f_1 + \frac{f_{1LL}}{2} - g_1 \right) , \\
    \frac{\bm{k}_\st^2}{2M^2} \left[ (f_{1T}^\perp + g_{1LT})^2 + (f_{1LT} + g_{1T} + h_{1LT})^2 \right] &\leq (f_1 - f_{1LL}) \left( f_1 + \frac{f_{1LL}}{2} + g_1 \right) , \\
    \frac{\bm{k}_\st^2}{2M^2} \left[ (f_{1T}^\perp - g_{1LT})^2 + (f_{1LT} - g_{1T} + h_{1LT})^2 \right] &\leq (f_1 - f_{1LL}) \left( f_1 + \frac{f_{1LL}}{2} - g_1 \right) , \\
    |h_{1TT}| &\leq \frac{1}{2} \left( f_1 + \frac{f_{1LL}}{2} + g_1 \right) , \\
    \frac{\bm{k}_\st^4}{2 M^4} \,|h_{1TT}^{\perp\perp}| &\leq f_1 + \frac{f_{1LL}}{2} - g_1 , \\
    \frac{\bm{k}_\st^4}{M^4} \left[ g_{1TT}^2 + (f_{1TT} - h_{1TT}^\perp)^2 \right] &\leq \left( f_1 + \frac{f_{1LL}}{2} + g_1 \right) \left( f_1 + \frac{f_{1LL}}{2} - g_1 \right) . \label{e:bound9}
\end{align}
These inequalities are relevant for the study of tensor polarized gluon TMDs at e.g.\ the EIC possibility at JLab (JLEIC)~\cite{Boer:2011fh,Abeyratne:2012ah,Abeyratne:2015pma} or COMPASS~\cite{Ball:2006zz} using tensor polarized deuterons. The proposed fixed-target experiment at LHC (AFTER@LHC)~\cite{Brodsky:2012vg} would also allow to investigate the gluon TMDs~\cite{Kikola:2017hnp}, in principle including ones related to tensor polarization.

Finally, we also include the bounds that apply to spin-$1/2$ hadrons. This case has been discussed already in~\cite{Mulders:2000sh}, however using a different notation and leaving the T-odd TMDs aside. The parametrization of the correlator for a spin-$1/2$ hadron is given by the sum of the terms~\eqref{e:term1}--\eqref{e:term3}. The density matrix is now parametrized in terms of the spin vector only and is a $2\times2$ matrix in hadron spin space. Using the decomposition in eq.~\eqref{e:average}, $G$ is a $4\times4$ matrix in gluon $\otimes$ hadron spin space and its explicit form (that \emph{does} contain the T-odd functions) is given in~\cite{Mulders:2000sh}. From that matrix we can extract the following bounds from its $2\times2$ principal minors:
\begin{align}
    |g_1| &\leq f_1 , \label{e:boundMR1} \\
    \frac{\bm{k}_\st^4}{4M^4} \left[ (h_{1L}^\perp)^2 + (h_1^\perp)^2 \right] &\leq (f_1 + g_1) (f_1 - g_1) , \\
    \frac{|\bm{k}_\st|}{M} \,|h_1| &\leq f_1 + g_1 , \\
    \frac{|\bm{k}_\st|^3}{2M^3} \,|h_{1T}^\perp| &\leq f_1 - g_1 , \\
    \frac{\bm{k}_\st^2}{M^2} \left[ (f_{1T}^\perp)^2 + g_{1T}^2 \right] &\leq (f_1 + g_1) (f_1 - g_1) . \label{e:boundMR5}
\end{align}
Upon omitting tensor polarization (and discarding all functions related to it) in bounds~\eqref{e:bound1}--\eqref{e:bound9}, which is mathematically equivalent to considering the spin-$1/2$ case, one obtains a set of bounds that is less strict (but consistent with) the bounds~\eqref{e:boundMR1}--\eqref{e:boundMR5}. In general, these less strict bounds can be sharpened upon considering the eigenvalues of higher-dimensional principal minors.

Note that in eq.~\eqref{relationTMD} we did not consider the process-dependent gauge link structure explicitly. In fact, the inequalities~\eqref{e:bound1}--\eqref{e:bound9} and~\eqref{e:boundMR1}--\eqref{e:boundMR5} do \emph{not} hold generally true for any correlator -- the matrix $G$ is positive semidefinite only for field combinations, including gauge links, that `factorize' into the form $O^\dagger(0) O(\xi)$. The simplest gauge link structures for which this holds are $[+,+]$, $[-,-]$, $[+,-]$, and $[-,+]$ (the plus and minus refer to the future- and past-pointing Wilson lines defined in eq.~\eqref{e:link}, and for the second entry Hermitian conjugation is implied). For the same links, one also has the constraint $f_1 \ge 0$ (in the spin-$1/2$ case this follows already from bound~\eqref{e:boundMR1}). The $[-,-]$ gauge link appears in processes with color flow annihilated within the initial state, such as the (gluonic) Drell-Yan process or Higgs production through gluon fusion ($gg \to h$)~\cite{Boer:2013fca,Echevarria:2015uaa}. The structure $[+,+]$, on the other hand, is related to color flow into the final state, which is the case for e.g.\ quark-antiquark pair production in semi-inclusive deep-inelastic scattering~\cite{Pisano:2013cya}. When color flow involves both initial and final states, the gauge links $[+,-]$ and $[-,+]$ appear, which is for instance the case in processes with $qg \to qg$ and $\bar{q}g \to \bar{q}g$ contributions respectively~\cite{Bomhof:2006dp}.

\subsection{Bounds on transverse momentum integrated functions}
We now turn to the transverse momentum integrated case, i.e.\ we will establish relations between the collinear PDFs appearing in eq.~\eqref{e:PDFpar}. Although this case was recently covered already in~\cite{Cotogno:2017mwy}, we include it here for completeness. The $3\times3$ blocks of the matrix $G$ in eq.~\eqref{Matrixsixbysix} are now given by:
\begin{equation*}
    A = \left( \begin{array}{ccc}
    {f_{1}}+\frac{{f_{1LL}}}{2}-{g_1} & 0 & 0 \\[5 pt]
    0 & {f_{1}}-{f_{1LL}} & 0  \\[5 pt]
    0 & 0 & {f_{1}}+\frac{{f_{1LL}}}{2}+{g_1} \\
    \end{array} \right) , \qquad B = \left( \begin{array}{ccc}
    0 & 0 & 0 \\[5 pt]
    0 & 0 & 0  \\[5 pt]
    -2 {h_{1TT}} & 0 & 0 \\
    \end{array} \right) ,
\end{equation*}
\begin{equation*}
    C = \left( \begin{array}{ccc}
    {f_{1}}+\frac{{f_{1LL}}}{2}+{g_1} & 0 & 0 \\[5 pt]
    0 & {f_{1}}-{f_{1LL}} & 0  \\[5 pt]
    0 & 0 & {f_{1}}+\frac{{f_{1LL}}}{2}-{g_1} \\
    \end{array} \right) ,
\end{equation*}
where we have suppressed the argument $(x)$ of the functions. From integration of eq.~\eqref{relationTMD} over transverse momentum, it follows that again $G$ is positive semidefinite. In contrast to the TMD case, we can easily diagonalize the full matrix and we obtain the following three bounds:
\begin{align}
    |g_1| &\leq f_1 + \frac{f_{1LL}}{2} , \label{Unp} \\
    f_{1LL} &\leq f_1 , \\
    |h_{1TT}| &\leq \frac{1}{2} \left( f_1 + \frac{f_{1LL}}{2} + g_1 \right) , \label{Soffer}
\end{align}
where, including also $f_1 \ge 0$, these inequalities hold for \emph{any} process as PDFs are universal. In the spin-$1/2$ case, one simply has the bound $|g_1| \leq f_1$.

Through the gluon structure function $\Delta(x)$ of~\cite{Jaffe:1989xy}, the PDF $h_{1TT}(x)$ is related to the double-helicity flip scattering amplitude in processes involving hadrons of spin $\ge$ 1. At the parton level, $h_{1TT}(x)$ represents the distribution of linearly polarized gluons in a transversely tensor polarized target sometimes referred to as `gluon transversity', a name we do not want to use as it may misleadingly suggest transverse polarization of the gluon. Recently in~\cite{Detmold:2016gpy} the first moment of a bound analogous to~\eqref{Soffer} was studied on the lattice in a $\phi$ meson ($s\bar{s}$). The bounds~\eqref{Unp}--\eqref{Soffer} will be relevant e.g.\ for the extraction of $\Delta(x)$, which has been proposed to occur at JLab using nitrogen targets~\cite{DetmoldJLab}, and which could also be achieved within the program of the EIC~\cite{Accardi:2012qut}.

\section{Positivity bounds on gluon TMDs at small x} \label{sec:4}
In this section we will consider bounds on the gluon TMDs in the small-$x$ kinematic region. The gluon-gluon TMD correlator simplifies greatly in the small-$x$ limit for the so-called dipole-type gauge link structure $[+,-]$, which has recently been shown in~\cite{Boer:2015pni,Boer:2016xqr}. More specifically,~\cite{Boer:2016xqr}
\begin{equation}
    \lim_{x\to0} \,\Gamma^{[+,-]\,ij}(x,\bm{k}_\st) = \frac{k_\st^i k_\st^j}{2\pi L} \;\Gamma_0^{[\Box]}(\bm{k}_\st) ,
    \label{e:smallxrel}
\end{equation}
where the so-called Wilson loop correlator appearing on the right-hand side is defined as
\begin{equation}
    \Gamma_0^{[\Box]}(\bm{k}_\st) \equiv \left. \int \frac{d^2\bm{\xi}_\st}{(2\pi)^2} \;e^{-i\bm{k}_\st\cdot\bm{\xi}_\st} \bra{P;S,T} \tr_c \big( U^{[\Box]} \big) \ket{P;S,T} \right|_{\xi\cd n = 0} , 
    \label{e:WLcorrelator}
\end{equation}
where $U^{[\Box]} \equiv U_{[0,\xi]}^{[+]} \,U_{[\xi,0]}^{[-]}$ is a rectangular Wilson loop with transverse extent $\bm{\xi}_\st$ and longitudinal dimension $L \equiv \int d\xi\cd P = 2\pi\,\delta(0)$. The parametrization of this correlator in terms of TMDs is given by~\cite{Boer:2016xqr}
\begin{equation}
    \Gamma_0^{[\Box]}(\bm{k}_\st) = \frac{\pi L}{M^2} \left[ e(\bm{k}_\st^2) + \frac{\epsilon_\st^{S_\st k_\st}}{M} \,e_T(\bm{k}_\st^2) + S_{LL} \,e_{LL}(\bm{k}_\st^2) \vphantom{\frac{k_\st^{\alpha\beta} {S_{TT}}_{\alpha\beta}}{M^2}} + \frac{\bm{k}_\st \cd \bm{S}_{LT}}{M} \,e_{LT}(\bm{k}_\st^2) + \frac{k_\st^{\alpha\beta} {S_{TT}}_{\alpha\beta}}{M^2} \,e_{TT}(\bm{k}_\st^2) \right] .
    \label{e:gamma0_tmds}
\end{equation}
In the small-$x$ limit and for the dipole-type gauge link structure, the gluon TMDs in eq.~\eqref{e:gamma_vp_tmds} reduce to the $e$-type Wilson loop TMDs according to eq.~\eqref{e:smallxrel}. The precise limits are found in table 1 of~\cite{Boer:2016xqr}. 

Also the Wilson loop correlator $\Gamma_0^{[\Box]}$ is a spin-averaged correlator, given by $\Gamma_0^{[\Box]} = \rho_{s^\prime s} \,G_{0\hspace{0.08cm}s s^\prime}^{[\Box]}$ (analogous to eq.~\eqref{e:average}). Inverting this relation, we find that $G_0^{[\Box]}$ is given by
\begin{equation}
    G_0^{[\Box]} = \frac{\pi L}{M^2} \left( \begin{array}{ccc}
    e + \frac{e_{LL}}{2} & \frac{e^{-i\phi} k }{\sqrt{2} M} \left(e_{LT} + ie_T\right) & \frac{e^{-2i\phi} k^2 }{M^2} \,e_{TT} \\[5 pt]
    \frac{e^{i\phi} k }{\sqrt{2} M}\left(e_{LT} - ie_T\right) & e - e_{LL} & -\frac{e^{-i\phi} k}{\sqrt{2} M} \left(e_{LT} - ie_T\right) \\[5 pt]
    \frac{e^{2i\phi} k^2 }{M^2} \,e_{TT}& -\frac{e^{i\phi} k }{\sqrt{2} M}\left(e_{LT} + ie_T\right) & e + \frac{e_{LL}}{2} \\
    \end{array} \right) ,
    \label{e:spinmatrix_loop}
\end{equation}
where we have suppressed the argument $(\bm{k}_\st^2)$ of the functions. In analogy to eq.~\eqref{relationTMD}, we can write the elements of $G_0^{[\Box]}$ in the following form:
\begin{align}
    G_{0\hspace{0.08cm}s s^\prime}^{[\Box]}(\bm{k}_\st) &\equiv \left. \int \frac{d^2\bm{\xi}_\st}{(2\pi)^2} \;e^{-i\bm{k}_\st\cdot\bm{\xi}_\st} \bra{P;s} \tr_c \big( U^{[\Box]} \big) \ket{P;s^\prime} \right|_{\xi\cd n=0} \nn \\
    &= \sum_m \,\tr_c \left( \bra{P_m} U^T_{[\infty,\bm{\infty}_\st;\infty,\bm{0}_\st]} \,U^n_{[\infty,\bm{0}_\st;-\infty,\bm{0}_\st]} \,U^T_{[-\infty,\bm{0}_\st;-\infty,\bm{\infty}_\st]} \ket{P;s}^* \right. \nn \\
    &\quad\, \left. \left. \times \bra{P_m} U^T_{[\infty,\bm{\infty}_\st;\infty,\bm{0}_\st]} \,U^n_{[\infty,\bm{0}_\st;-\infty,\bm{0}_\st]} \,U^T_{[-\infty,\bm{0}_\st;-\infty,\bm{\infty}_\st]} \ket{P;s^\prime} \right) \right|_{\xi\cd n = 0} \,\delta^{(2)}({\bm{P}_m}_\st + \bm{k}_\st) ,
    \label{e:posdef_loop}
\end{align}
where we inserted a complete set a momentum eigenstates $\{\ket{P_m}\}$ and we used that 
\begin{align}
    U^{[\Box]} &= U^n_{[-\infty,\bm{0}_\st;\infty,\bm{0}_\st]} \,U^T_{[\infty,\bm{0}_\st;\infty,\bm{\xi}_\st]} \,U^n_{[\infty,\bm{\xi}_\st;-\infty,\bm{\xi}_\st]} \,U^T_{[-\infty,\bm{\xi}_\st;\infty,\bm{0}_\st]} \nn \\
    &= \left( U^T_{[-\infty,\bm{\infty}_\st;-\infty,\bm{0}_\st]} \,U^n_{[-\infty,\bm{0}_\st;\infty,\bm{0}_\st]} \,U^T_{[\infty,\bm{0}_\st;\infty, \bm{\infty}_\st]} \right) \nn \\
    & \quad\; \times \left( U^T_{[-\infty,\bm{\infty}_\st;-\infty,\bm{\xi}_\st]} \,U^n_{[-\infty,\bm{\xi}_\st;\infty,\bm{\xi}_\st]} \,U^T_{[\infty,\bm{\xi}_\st;\infty,\bm{\infty}_\st]} \right)^\dagger .
    \label{e:ubox}
\end{align} 
From eq.~\eqref{e:posdef_loop} it follows that $G_0^{[\Box]}$ is positive semidefinite; in other words, its eigenvalues must be $\geq 0$. To establish bounds for the Wilson loop TMDs, we again restrict ourselves to two-dimensional principal minors. We obtain the following two inequalities:
\begin{align}
    \frac{\bm{k}_\st^2}{2M^2} \left( e_T^2 + e_{LT}^2 \right) &\leq (e - e_{LL}) \left( e + \frac{e_{LL}}{2} \right) , \label{e:ebound1} \\
    \frac{\bm{k}_\st^2}{M^2} \,|e_{TT}| &\leq e + \frac{e_{LL}}{2} . 
    \label{e:ebound2}
\end{align}
Applying the small-$x$ limit to the bounds~\eqref{e:bound1}--\eqref{e:bound9}, one indeed recovers the bounds~\eqref{e:ebound1} and~\eqref{e:ebound2}. Besides these two bounds, we also have $e \geq 0$ (this follows from eq.~\eqref{e:ubox}).

Let us finally also comment on the case of a spin-$1/2$ hadron. The parametrization of the Wilson loop correlator for spin-$1/2$ hadrons is given in terms of the two functions $e$ and $e_T$. Analogous to $G_0^{[\Box]}$ in eq.~\eqref{e:spinmatrix_loop}, we now obtain
\begin{equation}
    G_0^{[\Box]} = \frac{\pi L}{M^2} \left( \begin{array}{cc}
    e & \frac{ie^{-i\phi} k}{M} \,e_T \\[5 pt]
    -\frac{ie^{i\phi} k}{M} \,e_T & e \\
    \end{array} \right) ,
    \label{e:spin1/2}
\end{equation}
from which we can derive the following upper bound for $e_T$:
\begin{equation}
    \frac{|\bm{k}_\st|}{M} \,|e_T| \leq e .
    \label{spinHalfsmallx}
\end{equation}
Note that upon omitting tensor polarization and discarding all functions related to it (this is, in fact, mathematically equivalent to the reduction to a spin-$1/2$ description), the bounds~\eqref{e:ebound1} and~\eqref{e:ebound2} reduce to a bound that is consistent with but less strict than~\eqref{spinHalfsmallx}. Aside from diagonalizing higher-dimensional minors to sharpen the bounds, we can also obtain~\eqref{spinHalfsmallx} by applying the small-$x$ limit to the bounds for spin-$1/2$ hadrons given in~\eqref{e:boundMR1}--\eqref{e:boundMR5}.

\section{Discussion and conclusions} \label{sec:5}
We have studied positivity bounds on gluon-gluon correlators for hadrons including in particular the polarized spin-1 hadrons, thus looking at the unpolarized, vector polarized, and tensor polarized cases. The bounds have been derived using the fact that the correlators, even including gauge links that bridge the nonlocality, can be expressed as momentum densities. For both the TMD and collinear cases, we have obtained relations between the distribution functions that appear in the parametrization of the leading-twist gluon-gluon correlator. The bounds follow from the positive semidefiniteness of the correlation function, and rely primarily on the operator structure of the correlator and as such they can be considered as rigorous tests of QCD, provided the functions are compared at the same scales. 

For gluons, especially the small-$x$ region is important, which is why we have also studied the bounds for TMDs in the small-$x$ limit. To this end, we have exploited the main results of~\cite{Boer:2016xqr}, being the fact that for the dipole-type gauge link structure the gluon-gluon correlator simplifies to a correlator containing a single Wilson loop. The latter correlator can also be parametrized in terms of TMDs, for which we have found bounds as well.

The actual value of the established bounds reckon on the extraction of the functions from the cross-sections, which in turn relies on the validity of the leading-power expression of the cross-section in terms of the distribution functions, i.e.\ the absence of subleading powers. Furthermore, since we look at TMDs, one must worry about the process dependence coming from the fact that functions with different types of gauge links may be needed to describe a process at measured transverse momenta~\cite{Buffing:2013kca}. Another complication is that the dependence of the distributions on $\bm{k}_\st^2$ may require additional functions involving gluonic poles~\cite{Boer:2015kxa}. Assumptions on some of the TMDs and using approximations such as taking the large-$N_c$ limit, may thus be necessary. In addition to these points, one might also worry about the effects of QCD evolution on the validity of the bounds. In the collinear case, the Soffer bound involving three quark functions~\cite{Soffer:1994ww} has been shown to be preserved up to next-to-leading order accuracy~\cite{Vogelsang:1997ak,Bourrely:1997bx,Martin:1997rz}. However, to our knowledge, there are no studies yet on the stability of bounds under evolution concerning TMDs. The fact that the evolution kernel for TMDs is independent of spin~\cite{Echevarria:2014rua,Echevarria:2015uaa}, might suggest that in the appropriate $\bm{k}_\st$-regime where TMD factorization is valid, positivity bounds are respected also in this case. However, the latter could depend on the specific implementation of TMD evolution. This topic remains open to further investigation.

The results in this paper may be relevant for proposed experiments at JLab and a future EIC involving tensor polarized targets. In practical situations, however, the bounds will, rather than serving as a test of QCD, often be more useful as a check in models or lattice calculations, or as a way to obtain an order of magnitude estimate of TMDs. The latter is commonly done by saturating the bounds. These estimates for the functions then can be used, for instance, as input for an estimate of measurements of particular azimuthal and spin asymmetries.

\begin{acknowledgements}
We thank Miguel Garc\'ia Echevarr\'ia and Andrea Signori for fruitful discussions, and Alessandro Bacchetta and Dani\"el Boer for useful feedback on the manuscript. This research is part of the European Research Council (ERC) under the program QWORK (contract no. 320389).
\end{acknowledgements}

\bibliographystyle{JHEP}
\bibliography{references}
\addcontentsline{toc}{section}{References}

\end{document}